\documentstyle[sprocl]{article}

\def\Journal#1#2#3#4{{#1} {\bf #2}, #3 (#4)}

\def\NP{{\em Nucl.\ Phys.}}

\def\PRL{\em Phys.\ Rev.\ Lett.}
\def\PRC{{\em Phys.\ Rev.}~C}

\def\be{\begin{equation}}
\def\ee{\end{equation}}
\def\bea{\begin{eqnarray}}
\def\eea{\end{eqnarray}}
\def\gapproxeq{\raisebox{-.5ex}{$\,\stackrel{>}{\scriptstyle
\sim}\,$}}

\def\etal{{\em et al.}}

\begin{document}

\title{COMPRESSION AND EXPANSION IN CENTRAL
COLLISIONS\footnote{Talk given at the Int.\ Research Workshop
on ``Heavy Ion Physics at Low, Intermediate, and Relativistic
Energies with 4$\pi$ Detectors", Poiana Brasov, Romania, 7-14
October, 1996.}}

\author{ P.~DANIELEWICZ }

\address{
      National Superconducting Cyclotron Laboratory\\
       and Department of Physics and Astronomy\\
       Michigan State University, East Lansing, MI 48824-1321, USA}


\maketitle\abstracts{
Dynamics of central collisions of heavy nuclei in the energy
range from few tens of MeV/nucleon to a~couple of GeV/nucleon
is discussed.   As~the
beam energy increases and/or the impact parameter decreases,
the maximum
compression increases.  It~is argued that the hydrodynamic
behavior of matter sets in the vicinity of balance energy.
At~higher energies shock
fronts are observed to form within head-on reaction
simulations,
perpendicular to beam axis and separating hot compressed matter
from cold.
In the semicentral
reactions a weak
tangential discontinuity develops in-between these fronts.  The
hot compressed matter exposed to the vacuum in directions
parallel to the shock fronts  begins to expand collectively
into these directions.  The expansion affects particle angular
distributions and mean energy components and further shapes of
spectra and mean energies of particles emitted into any one
direction.  The~variation of slopes and the relative yields
measured
within the FOPI collaboration are in a~general agreement with
the results of simulations.
As~to the FOPI data on stopping, they
are consistent with the
preference for transverse over the longitudinal motion in the
head-on Au + Au
collisions.  Unfortunately, though, the data cannot be used to
decide directly on that preference due to acceptance cuts.
Tied to the spatial and temporal changes in the reactions are
changes in the entropy per nucleon.
}

\section{Introduction}

In my talk, I shall discuss the dynamics of central heavy-ion
reactions in the energy range from few tens of MeV to a couple
of GeV per nucleon.  I~shall use simple
estimates, data, and transport reaction simulations, to~reach
conclusions.  When moving from low to high energies, from light
to heavy systems, or from high to low impact parameters,
I~will
argue, the~onset of hydrodynamic behavior in collisions is
demarked by the balance energy.  Transport simulations indicate
that in the high-energy head-on collisions of nuclei, in~the
hydrodynamic regime, nuclear shock waves develop,
perpendicular to the beam axis, separating hot compressed
nuclear matter from cold matter.  With time, the~compressed
matter exposed to the vacuum acquires collective motion
giving rise to a~characteristic behavior of transverse cm
spectra of  different particles with mass.  At~finite impact
parameters a~so-called weak discontinuity develops in-between
the shocks.  Both the motion associated with the discontinuity
and the expansion contribute to the sideward flow.  Recently
FOPI data became available on stopping and on expansion in
central collisions.  I~shall examine to what extent these data
are consistent with the above scenario.  Finally, I~shall
discuss the production of entropy in collisions and
conclude.

\section{Onset of Hydrodynamic Behavior: Balance Energy}

The~hydrodynamic behavior requires a~local equilibrium and,
thus, a~small mean free path,
$\lambda \sim 1/(\rho \, \sigma_{\rm NN}) \sim 2$~fm, compared
to the size
of a~system, $\lambda / R \ll 1$.  Here $\rho$ is density and
the effects of Pauli principle are ignored.  The~condition
implies that
the~hydrodynamic behavior is more likely in the interactions of
heavy than of light nuclei, and in the more central than
in peripheral collisions.  Adding to this consideration,
the effects of the Pauli, lengthening the mean
free path, can
slow down the equilibration required for the hydrodynamic
behavior, at low beam energies.
At~moderate energies the momentum dependence of nuclear mean
field
may cause a~reduction in the number of NN collisions affecting
equilibration.

At~low energies,
nuclei colliding at moderate to high impact
parameters are expected to
clutch and rotate in a~binary system, emitting particles to
negative angles.  At~high energies, high pressure
should develop in the region between the nuclei, expelling
nucleons and the residual
nuclei to positive angles.  For~emission quantified in
terms of the average momentum in the
reaction
plane as a~function of rapidity
$\langle p^x/A
\rangle (y) $,
the~slope of the dependence on rapidity
should then change sign,
when going from low to high bombarding
energies, at the so-called balance energy.
I~will argue that,
whether in the bombarding energy or in the impact parameter,
the~balance energy demarks the onset of the hydrodynamic
behavior.

\begin{figure}
  \vspace*{7.3cm}
  \includegraphics{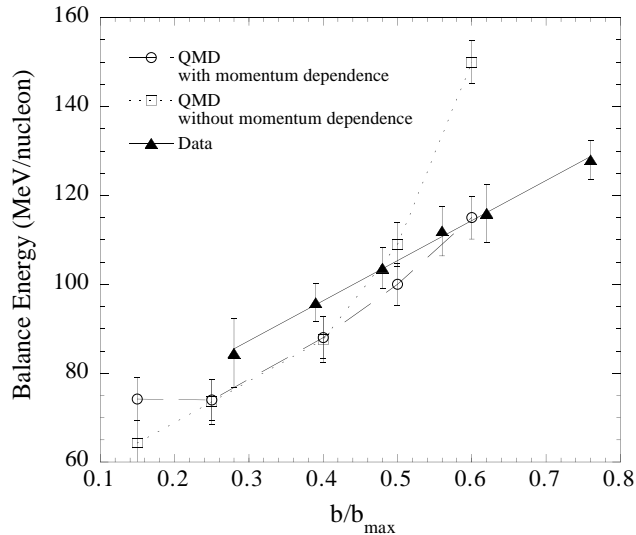}
\caption{ Balance energy
in the $^{40}$Ar + $^{45}$Sc
system
as a function of reduced impact
parameter.  Triangles and points
represent, respectively, data of Ref.~\protect\citelow{pak96}
and calculations of Ref.~\protect\citelow{sof95}
}
\label{balb}
\end{figure}

\begin{figure}
  \vspace*{7.6cm}
  \includegraphics{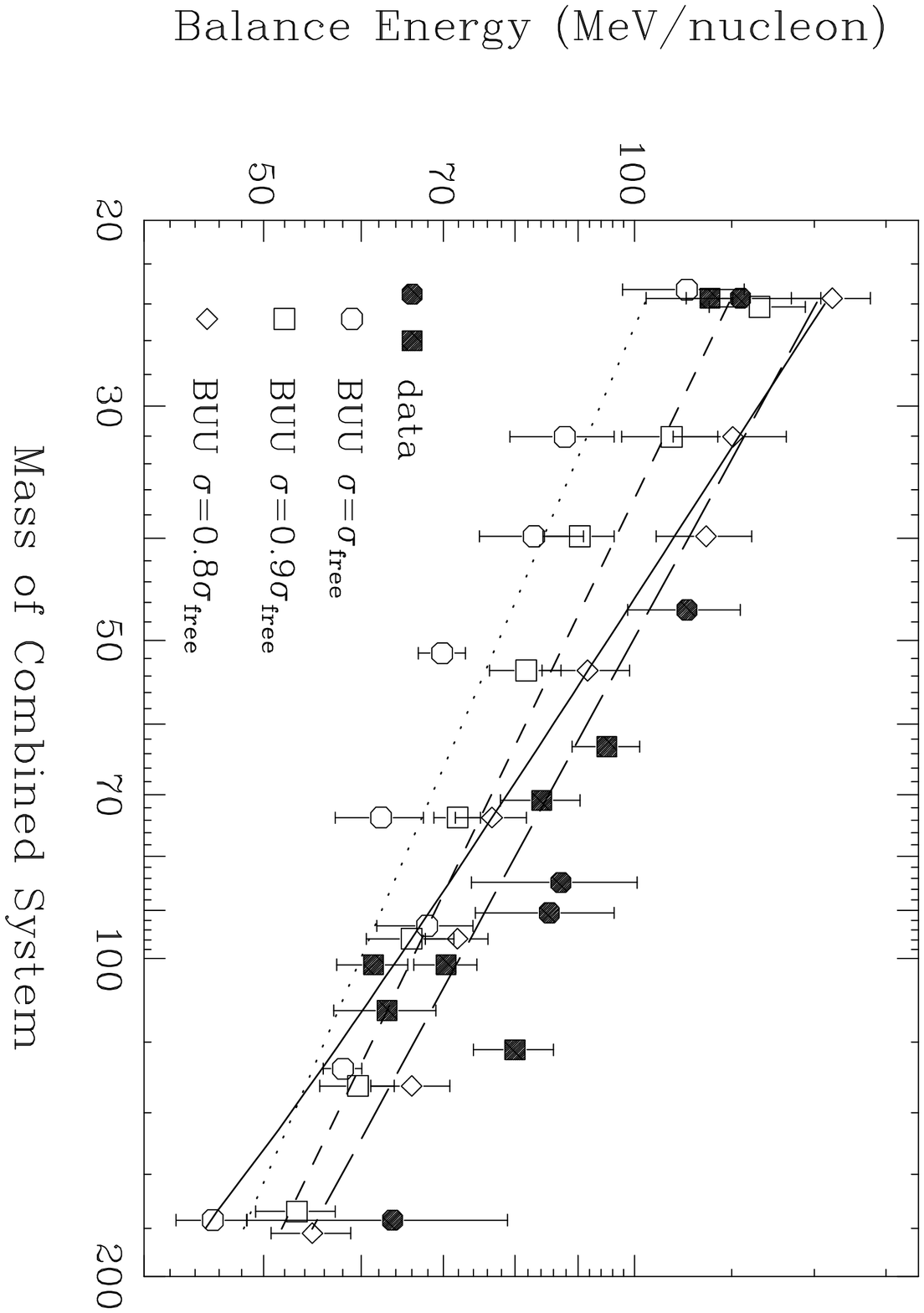}
\caption{ Balance energy in nearly symmetric systems as
a~function of total mass.  Filled squares and circles
represent, respectively, data of Ref.~\protect\citelow{wes93}
and~\protect\citelow{but95}  Open symbols represent
transport-model
calculations of Ref.~\protect\citelow{kla93}  The~dashed and
dotted lines indicate the general trends of the
results.\protect\cite{kla93}  The~solid line represents an
estimate according to Eq.~(\protect\ref{EBAL}).
}
\label{balE}
\end{figure}

By~investigating the magnitude of the flow,
$d \langle p^x/A
\rangle / dy $ at midrapidity, the~balance energy
has been studied~\cite{pak96} as a~function of the
reduced impact parameter~$\hat{b} = b/b_{max}$,
in~the system $^{40}$Ar~+~$^{45}$Sc.  The~energy was
observed to rise with the rise in~$\hat{b}$,
see Fig.~\ref{balb}.
The~measurement is
consistent with the hydrodynamic behavior setting earlier in
the more central than in the more peripheral collisions
as the beam energy increases (see also Ref.~\citelow{but95}).
The~dependence of the balance energy on total mass for
symmetric
systems is shown in Fig.~\ref{balE}.
The~balance energy is
higher for light systems and lower for heavy systems,
consistent with the hydrodynamic behavior setting earlier in
the heavy than in the light systems.  The~balance energy
calculated
with the transport model~\cite{kla93} strongly depends on NN
cross sections; better agreement with the data is obtained for
reduced than for free-space cross-sections.

In~the measurements
such as e.g.,\citelow{pak96,wes93,but95}
there has been no control on the direction of the
reaction plane.  Thus only a~suppression of the slope of
$\langle p^x/A
\rangle (y) $ at midrapidity could be observed as
a~function of the beam energy.
Principally, one~might imagine that at the deduced balance
energy only a~suppression of the slope takes place without
a~change in sign.  Recently in the
experiment,\citelow{lem96} though, the~polarization of
$\gamma$ rays
from a~residual target nucleus has been measured, in~the
$^{14}$N
+ $^{154}$Sm reaction at 35 and 100~MeV/nucleon, in~coincidence
with the emission of energetic $\alpha$ particles at
$\theta_{lab}=30^{\circ}$.  For~intermediate
reduced impact parameters~$\hat{b}$, the~$\gamma$ polarization
was found to change {\em sign} in-between 35 and 100
MeV/nucleon, in~a~manner
consistent with the change of {\em sign} of the flow, from
negative (attractive) to positive (repulsive).
For~high~$\hat{b}$, the~sign of the
polarization was found such as for an~attractive flow at either
of the beam energies.

Let us try to understand the mass dependence of the balance energy
in some detail; we~shall consider a~symmetric reacting system.
As~the nuclei come into contact, the~forces
that act on each include the force due to pressure in the neck
region with equilibrium and dissipative terms, the~proximity
force acting within the circumference of the neck, and the
Coulomb force,
\be
{\bf F}={\bf F}_{\rm eq+dis} + {\bf F}_{\rm prox}+{\bf
F}_{\rm Coul} \, .
\ee
The pressure within the neck region is
\be
P = P_{\rm eq} + {2 \over 3} \, \eta \, \nabla {\bf v} - 2  d
\, \rho \, \nabla^2 \left( {\rho \over \rho_0} \right) \, ,
\label{pressure}
\ee
where the first term on the r.h.s.~is equilibrium pressure,
next is a~portion of the viscosity tensor that
contributes to transverse force, and final term
is the finite-range correction.  The~factor $\eta$ in
the viscous term is viscosity coefficient.  The~proximity
force may be estimated either from the finite-range term
in~(\ref{pressure}) or by considering the surface energy within
the~system, obtaining
\be
F_{\rm prox} = -2 \pi \, R \, \sigma \, ,
\ee
where $\sigma \simeq 1.0$~MeV/fm$^2$ is the surface tension and
$R$
is the nuclear radius.  The~proximity force is proportional to
neck
circumference.  The~Coulomb force may be estimated as
$F_{\rm Coul} = Z_1 \, Z_2 \, e^2 /(R_1 + R_2)^2 \simeq 0.016
\,
A^{4/3}$~MeV/fm.  Given that no compression is expected at high
impact parameters,\cite{dan95} the~equilibrium pressure can be
estimated as~$P_{\rm eq} = 2\rho_0 \, E_{\rm cm} = \rho_0 \,
E_{\rm lab}/6$.
Finally, the~dissipative viscous correction to pressure may be
estimated as $P_{\rm dis} = - (2/3) \, (\rho_0 \, p_F \,
\lambda
/3) \, v_{\rm lab}/R$  (an~additional, nominally dissipative,
correction may arise from
the momentum dependence of the mean field).  At~the balance
energy the net force acting on any of the nuclei vanishes, i.e.
\be
0 = F_{\rm prox} + F_{\rm Coul} + \pi \, R^2 \, (P_{\rm eq} +
P_{\rm dis}) \, , \label{zero}
\ee
where $\pi \, R^2$ has been used as an estimate for the neck
area.  Solution of Eq.~(\ref{zero}) gives
\bea
\nonumber
E_{\rm bal} & \simeq & 7 \epsilon_F \, \left( {\lambda
\over R} \right)^2 + {24 \, \sigma \over R \, \rho_0}
- 0.21 \, A^{2/3} \\
& \simeq & {740 \over A^{2/3} } + { 160 \over A^{1/3} } - 0.21
\, A^{2/3} \, ,
\label{EBAL}
\eea
where $\epsilon_F$ is Fermi energy and the last result is given
in~MeV ($A$ is the {\em total} mass of the system).
The~crude estimate~(\ref{EBAL}) is shown with a~solid line
in~Fig.~(\ref{balE}).  Quantitative result allows to conclude the
following.  The~nonequilibrium correction in the pressure is
very important; if~this correction were omitted the balance
energy would turn out to be 4 times too low compared to data.
Observed large balance energies demonstrate that little
compression is reached in periphery; if~one assumed that all
c.m.~energy per nucleon were used up in a~compression,
the~balance energy from~(\ref{zero}) would underestimate
data by an~order of magnitude.
The~sensitivity to elementary cross sections for light
nuclei,
$E_{bal} \propto \sigma_{\rm NN}^{-2}$ with
$\sigma_{\rm nn} < \sigma_{\rm np}$,
seems to explain the
rise of
balance energy with the rise in the isospin asymmetry found in
the transport
calculations;~\cite{bao96} the~weakening of the mean field
should act to
reduce rather than to increase the balance energy.

\section{Head-On Reaction Simulations}

\begin{figure}
  \vspace*{10.8cm}
  \includegraphics{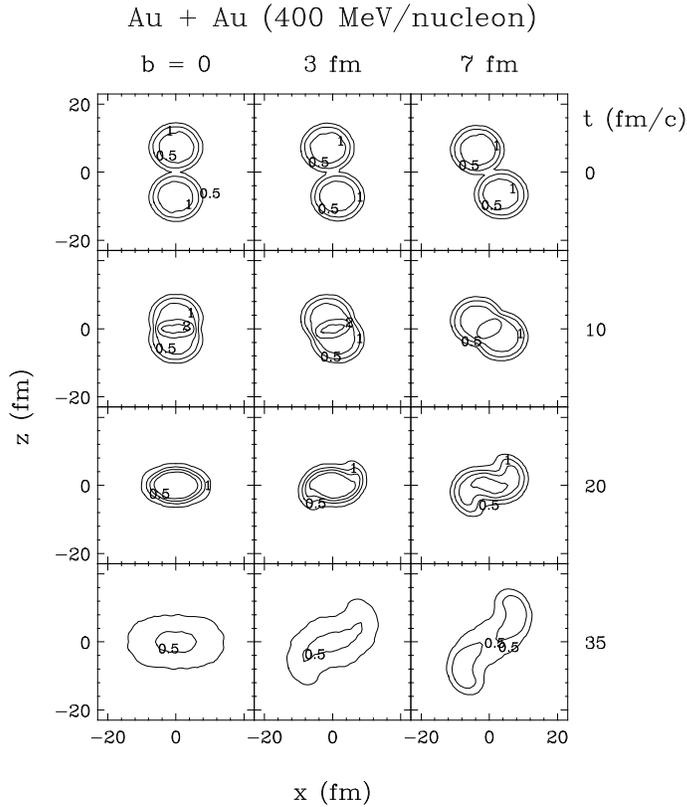}
\caption{
Contour plots of baryon density in the reaction plane in Au +
Au collisions at 400 MeV/nucleon.  The displayed contour lines
are for the densities $\rho/\rho_0$ = 0.1, 0.5, 1, 1.5, and~2.
}
\label{contours}
\end{figure}

\begin{figure}
  \vspace*{8.0cm}
  \includegraphics{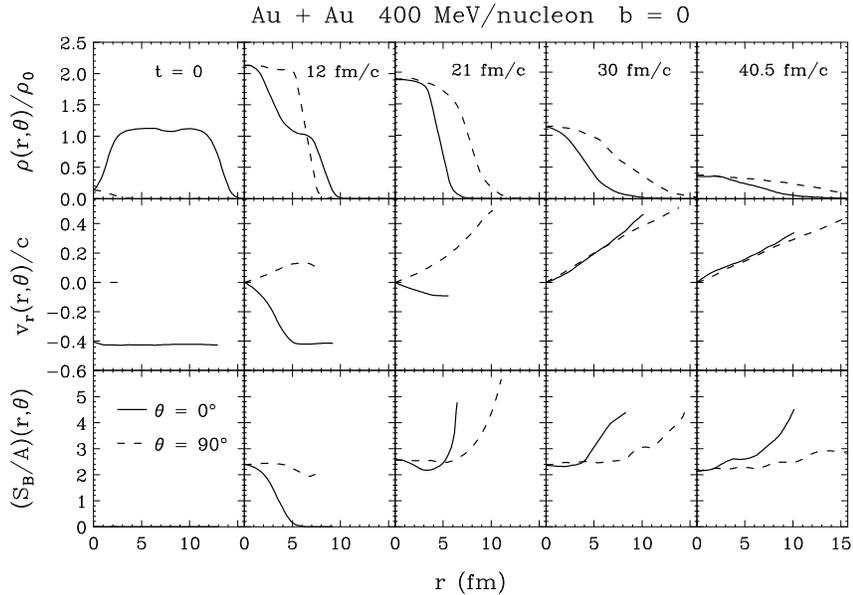}
\caption{
Baryon density (top panels), radial velocity (center
panels), and entropy per baryon along (solid lines) and
perpendicular (dashed lines)
to the beam axis at different indicated times, in the $b = 0$
collision at a beam energy of 400 MeV/nucleon.
}
\label{profau}
\end{figure}

Dynamics of nuclear reactions above the balance energy has been
studied in some detail following simulations based on the
Boltzmann equation with a~density and isospin dependent mean
field, free NN cross-sections, and composite production,
in~Ref.~\citelow{dan95}
and elsewhere.
It~is possible to break the dynamics
into several stages.
A~400 MeV/nucleon Au + Au system at $b = 0$, for which
baryon-density contour-plots are shown in Fig.~\ref{contours},
will serve to illustrate points.
Unless otherwise indicated the results of simulations will
correspond to a~stiff equation of state with
an~incompressibility of~$K= 380$~MeV.
Following an initial interpenetration of projectile and target
densities, the $NN$ collisions begin to thermalize matter in
the overlap region making the momentum distribution there
centered
at zero momentum in the c.m.s.  The density in the overlap
region rises above normal and a disk of excited
and compressed matter forms at the center of the system.  More
and more matter dives into the region with compressed matter
that begins to expand in transverse directions.  At
late stages, when the whole matter is excited, transverse
expansion predominates.
A further view at the situation in reaction at
$b=0$
is given in Fig.~\ref{profau}.

\begin{figure}
  \vspace*{10.0cm}
  \includegraphics{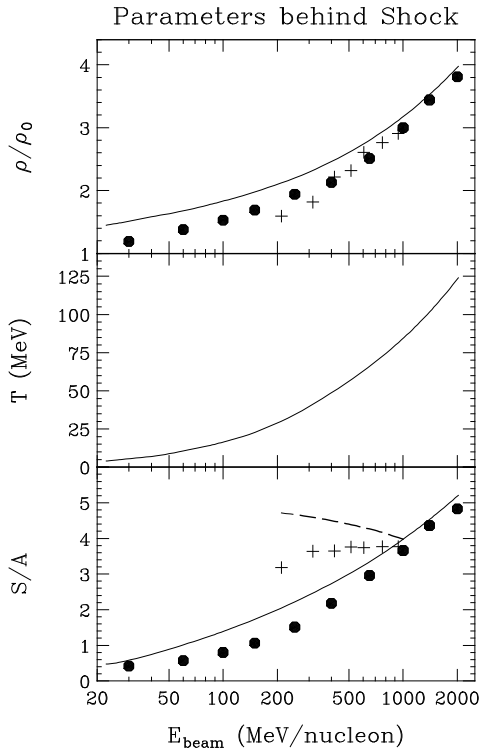} \caption{
Solid lines show the baryon density (top panel), temperature
(center panel), and entropy
per baryon (bottom panel) expected behind a developed shock
front at $b=0$, as a function of beam energy,
from solving the Rankine-Hugoniot equation.  Filled circles
indicate the maximum
density (top panel) and entropy per baryon at maximum
density (bottom panel) from
the simulations of $b=0$ Au + Au collisions at different beam
energies.
Crosses in the top panel indicate the maximum density
from the
simulations of 1~GeV/nucleon Au + Au collisions at $b$ = 3, 6,
8, 9, 10, 11, and
12 fm, plotted against the beam energy scaled down by a factor
equal nonrelativistically to $(1 - b^2/4R^2)$,
cf.~Eq.~(\protect\ref{efen}). }
\label{rankhu}
\end{figure}

If a nuclear system were very
large, then it would be necessarily ruled by laws of
hydrodynamics.
In the initial state of a head-on reaction, a discontinuity
exists in the velocity
field at the contact area of two nuclei, with the velocities in
the nuclei directed opposite.
Within
hydrodynamics, such an initial discontinuity
would have to break at finite times into two shock
fronts
travelling in opposite directions into the projectile and
target.
To assess whether the interfaces between normal and compressed
matter, cf.~Figs.~\ref{contours} and~\ref{profau}, may be
interpreted as the shock fronts, one can
ask whether the state of the matter at the center is such as
expected behind a~developed front.  Properties of the matter
behind
a~front can be determined, in a~straightforward
manner, from the
conservation of hydrodynamic fluxes of baryon number, momentum,
and energy.
Figure \ref{rankhu}
displays some of these properties, baryon density, temperature,
and entropy per baryon, together with a maximum density and
entropy
per baryon at a maximum density, from simulations.  The latter
quantities
generally follow the trends anticipated from hydrodynamics.
The shocks are nearly completely
developed in the high-energy simulations of
the head-on reactions of heavy nuclei.

Let us now turn attention to the expansion.  The~hot
matter in-between
the shocks is exposed in transverse directions to
vacuum and, as the pressure of the hot matter is finite and
that of
vacuum is zero, the matter begins to expand collectively into
transverse directions.
The features of this
process may be understood
at a {\em qualitative} level
in terms
of the self-similar
cylindrically-symmetric hydrodynamic expansion.
In the self-similar expansion, the velocity
is proportional
to the distance from symmetry axis, $v = {\cal F} (t)\,r$,
which
roughly holds in the direction of 90$^{\circ}$ in
the Au~+~Au collision at 400 MeV/nucleon, at~the center of the
system, see Fig.~\ref{profau}.  On~solving, in a~near-analytic
manner, the hydrodynamic equations of motion for the
expansion,\cite{dan95}
one arrives at the expectation that the matter should
accelerate collectively
in the transverse direction for a~time $R/c_s$, where $c_s$ is
the speed of sound.  This time is such as necessary for
a~signal
to propagate into the interior.  It is generally consistent,
$R/c_s
\sim 20$ fm/c, with the time during which collective transverse
energy, calculated from the local collective velocities
\begin{equation}
E^{\perp }_{\rm coll} =  \int d{\bf r} d{\bf p} \, f \,
m_N (v^{\perp })^2/2  ,
\label{Ecol}
\end{equation}
rises in the simulation displayed in Figs.~\ref{contours}
and~\ref{profau}.

After shocks reach the vacuum after a time
$\sim (2R/v_0)(\rho_1 -
\rho_0)/\rho_1 \sim 15$ fm/c, where
$v_0$ is the initial
velocity in the c.m.~and $\rho_1$ is  the density in
the shocked
region, an expansion
{\em along} the beam axis sets in, see~Fig.~\ref{profau}.
However, as the expansion in transverse
directions is already in progress and matter becomes
decompressed,
the expansion in longitudinal direction does not acquire same
strength.

While definite features
of reaction dynamics
appear to be consistent with a hydrodynamic behavior of the
matter, there
are also important differences.
For example, in a system continuing to expand
hydrodynamically
temperatures would have dropped to zero and {\em all} kinetic
energy would have got converted into collective energy of
expansion.  For that collisions would need
to continue down to very low densities.  The transverse
collective energy from~(\ref{Ecol}) freezes out at about
40~MeV/nucleon.  With about 10 MeV/nucleon of
longitudinal collective energy, one finds that for nucleons the
net collective energy constitutes about 50\% of the kinetic
energy in the c.m.~for a~400~MeV/nucleon reaction.

\section{Semicentral Reactions}

\begin{figure}
  \vspace*{5.4cm}
  \includegraphics{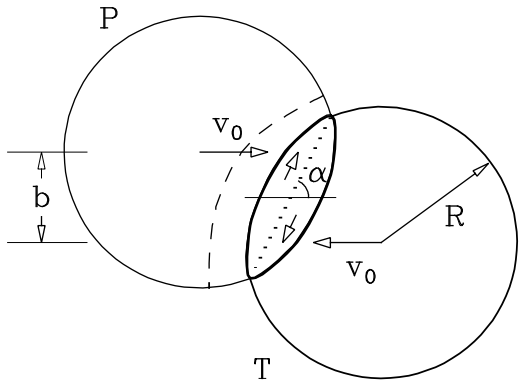} \caption{
The initial discontinuity between the projectile and target
velocities breaks at a finite $b$ into
two shock
fronts propagating into the projectile $P$ and target $T$
(thick solid lines), and a weak tangential discontinuity
in-between (dotted line).  Shock-front position {\em within}
projectile at a later time is indicated with a dashed line.
}
\label{sketch}
\end{figure}

In the initial state, an angle $\alpha$ of inclination of the
plane of discontinuity in
velocity, relative to the beam axis,  is given
nonrelativistically by
\begin{equation}
\cos \alpha \approx {b \over 2 R} .
\label{alpha}
\end{equation}
While at $b=0$ velocities are normal to the plane of
discontinuity, the velocities have finite tangential components
relative to that plane at finite $b$.  Such components are
continuous across the shocks which detach from the initial
plane
of discontinuity and, with tangential components being directed
opposite for the matter from a~projectile and from a~target, at
the center of the system
a so-called tangential discontinuity develops, see the sketch
in Fig.~\ref{sketch}.
Density, pressure and entropy are continuous across that
discontinuity.  Only the tangential velocity component changes.

As only the normal velocity component drops to zero across
shocks, the shocks are weaker at a finite $b$ than in a head-on
reaction.  With the normal component being equal
to $v_0 \sin
\alpha$, the effective c.m.~kinetic energy for a shock at a
finite $b$ becomes
\begin{equation}
{E_{\rm lab} \over 4} \cdot \sin^2 \alpha =
{E_{\rm lab} \over 4} \, \left(1 - \cos^2 \alpha \right) =
{E_{\rm lab} \over 4} \, \left(1 - {b^2 \over 4 R^2} \right),
\label{efen}
\end{equation}
rather than $E_{lab}/ 4$ in the nonrelativistic
approximation.  In consequence,
the density behind a shock at a finite $b$
should be lower than at $b=0$ and it
should, in fact, coincide with the density
in a $b=0$ reaction at beam energy reduced by a
factor $(1-b^2/4R^2)$.  That is tested in the upper panel
of Fig.~\ref{rankhu}.  In the 1~GeV/nucleon reaction the
density appears to follow the above
expectation up to very high impact parameters.

\begin{figure}
  \vspace*{9.5cm}
  \includegraphics{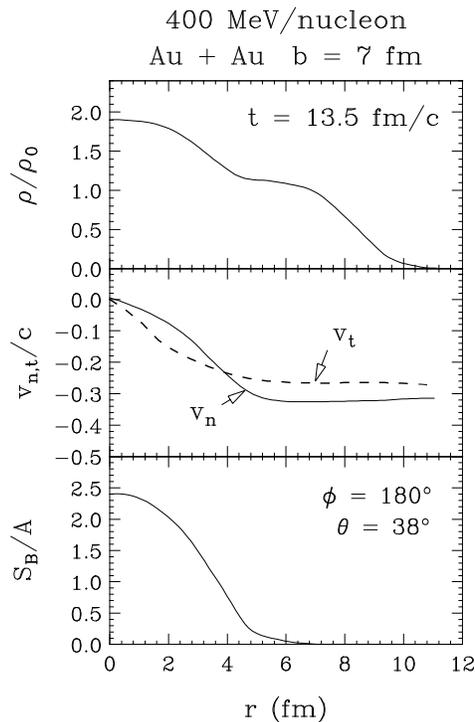} \caption{
Baryon density $\rho$ (top panel),
normal and tangential
velocity-components, $v_n$ and $v_t$, to the
discontinuities in nuclear-matter (center panel),
and entropy per baryon (bottom panel) as a function of the
distance from the center of a 400 MeV/nucleon Au + Au system at
$b = 7$ fm, along the normal to the discontinuities.  At the
given time $t = 13.5$ fm/c the discontinuities (at their
centers) are inclined
at an angle $\alpha \sim 52^{\circ}$ relative to the beam axis.
 }
\label{tang}
\end{figure}

While a shock width generally does not change with time, if
conditions at the two sides of a shock wave stay
the same, the width of a weak discontinuity increases with
time.  That this is the case, may be seen by adapting the
Navier-Stokes equations to the conditions in the vicinity of a
weak discontinuity and by observing that the equations reduce
there
to a diffusion equation for a tangential velocity component.
The diffusion coefficient is equal
to the kinetic viscosity coefficient
 $\nu = \eta/(m_N \rho)$, where
$\eta$ is viscosity.  With an initial condition for the
tangential velocity in the ideal-fluid limit in the form $v_t =
v_t^0 \, \epsilon(r)$, the solution to the diffusion equation
becomes $v_t(r,t) = v_t^0
\, \mbox{erf}\left({r / 2 \sqrt{\nu t}} \right)$, where $r$
is the distance along a line perpendicular to the
discontinuity,
i.e.~the width increases as $t^{1/2}$.  If a nuclear system
were very large, then the two shock waves and a weak
discontinuity would well separate from each other, as the
distance between the shocks increases as~$t$.  In actual
simulations, though, the tails of discontinuities overlap,
cf.~Fig.~\ref{tang}.
In that figure, the center of the shock is at a~position $r
\sim 3.3$ fm, where
density rises to the mean of that in normal matter and that
at the center of the system, where, further, entropy per
nucleon rises to a half of its maximum value, and where
magnitude of the normal velocity component drops to a half of
the asymptotic value.  The center of the tangential
discontinuity in the figure is, on the other hand, at $r=0$,
where tangential velocity
component vanishes.  The width of the discontinuity is
characterized by $\delta
\sim 1.7$ fm which agrees well with $\delta = \sqrt{\eta t
/ m_N \rho}$ expected from the simple formula, given above, for
the space and time behavior of the tangential velocity.

\begin{figure}
  \vspace*{7.6cm}
  \includegraphics{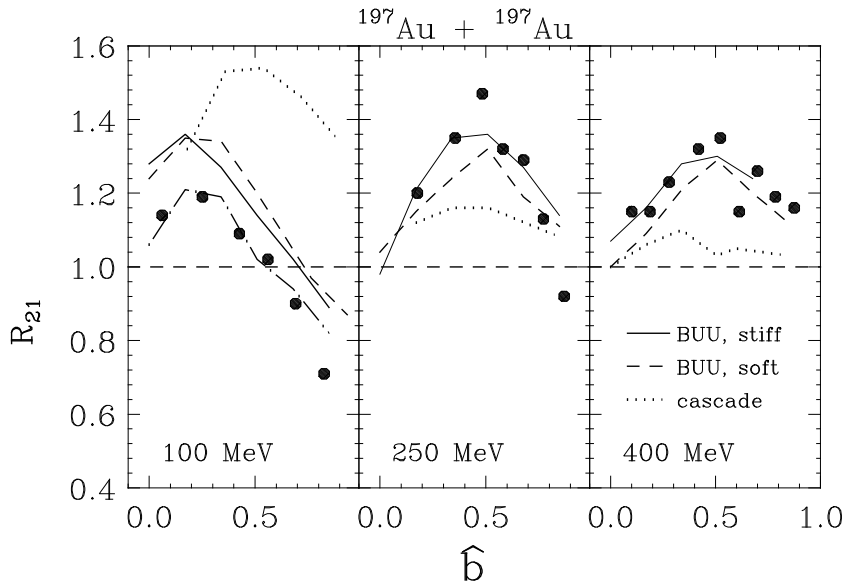}
\caption{
Ratio of the out-of-reaction-plane
mean energy-component to the lower in-plane eigenvalue of the
kinetic energy tensor, as a function of reduced
impact parameter, in Au~+~Au collisions at beam energies
indicated in the panels.  Circles represent the measurements of
Ref.~\protect\cite{tsa96}.  Solid and dashed lines represent
the results of the transport-model calculations with mean
fields yielding a
stiff and soft equation of state, respectively.  The~dotted
lines represent the results of the cascade model.
 }
\label{r21}
\end{figure}

Let us start to confront the findings with some data.
Just as at $b=0$ in a~heavy system, an~expansion develops at a
finite $b$ in-between
the shock waves, as an~effect of the exposure of the hot matter
to vacuum.
Anisotropy in the
expansion, with regard to the
direction perpendicular to the reaction plane and the direction
of the shock motion, results from the delay in
the start of expansion in the latter direction, as in
the case $b=0$.  The anisotropy (squeeze-out) may be quantified
with a ratio of
the eigenvalue of the kinetic energy tensor associated with the
direction out of the reaction plane, to the smaller of the
eigenvalues associated with the directions within the
plane,
\begin{equation}
R_{21} = {\langle E_2 \rangle \over \langle E_1 \rangle}.
\label{R21}
\end{equation}
As $b \rightarrow 0$ in a heavy system (but not in light),
the~ratio $R_{21}$ tends to $\langle
E^{\perp } \rangle /2 \langle E^{\parallel} \rangle $.
The Au + Au data for the ratio \cite{tsa96} are compared to the
results
of calculations, corrected for the experimental inefficiencies,
in Fig.~\ref{r21}.
At~250~MeV/nucleon and 400~MeV/nucleon
a~degree of agreement between the data and the calculations is
observed.  The~inefficiencies reduce the sensitivity of the
ratio to the nuclear compressibility.

\begin{figure}
  \vspace*{8.5cm}
  \includegraphics{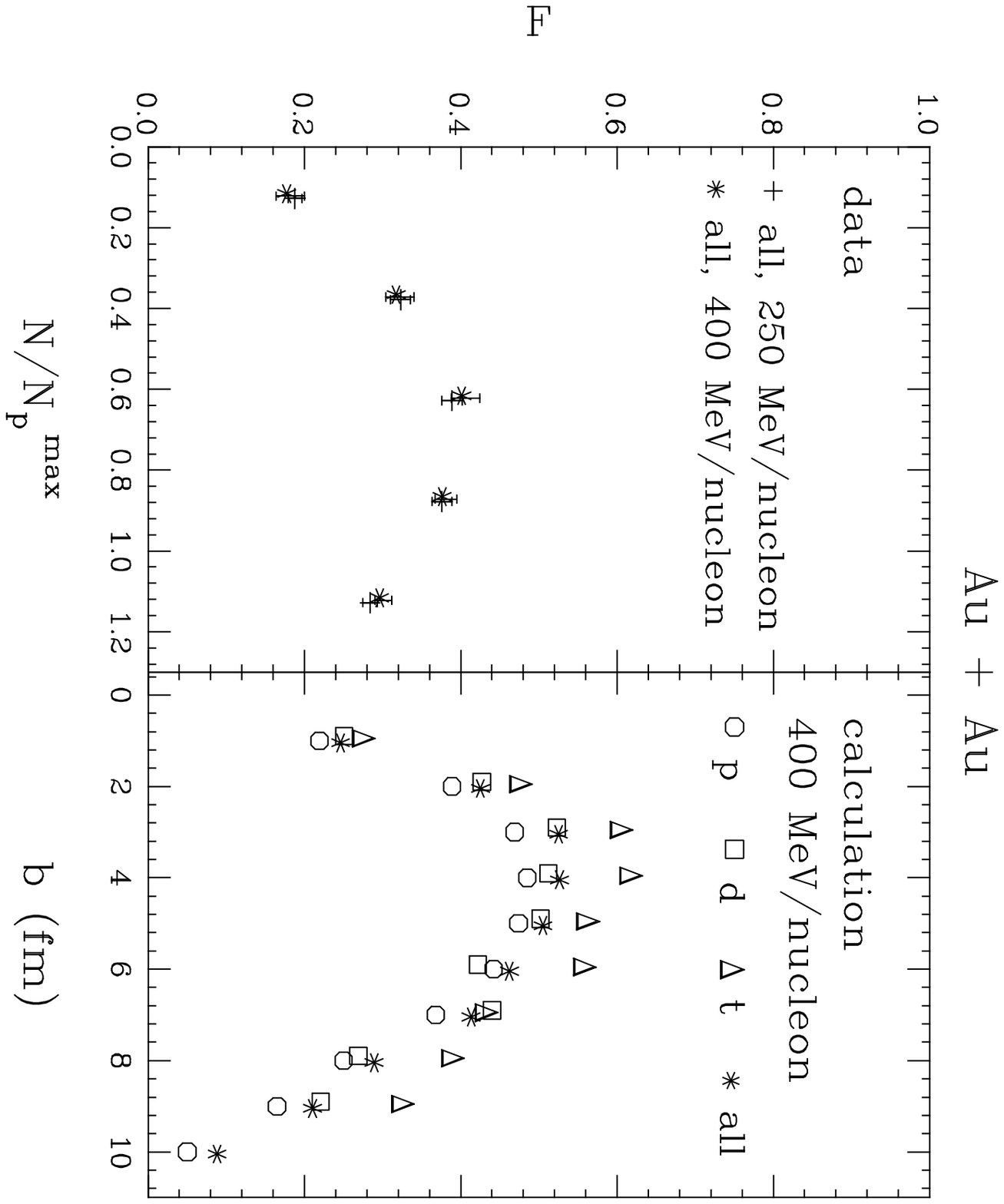}
\caption{
Left panel shows the flow parameter~$F$
in Au + Au
reactions at 250 and 400 MeV/nucleons, from the measurements of
Ref.~\protect\cite{gus88}, as a function of the reduced
participant
proton multiplicity.  Right panel shows the flow parameter from
the simulations of Au + Au reactions at 400 MeV/nucleon, as a
function of impact parameter.
 }
\label{slope}
\end{figure}

With regard to the directions within the reaction plane, both
the tangential motion behind the shock waves
for finite
impact parameters, and the anisotropy in expansion, contribute
to the
sideward deflection quantified usually in terms of the slope of
mean momentum within the reaction plane at midrapidity,
$F =  {d \langle p^x/m \rangle / dy}$.
Measured and calculated slopes in the 400~MeV/nucleon Au + Au
reaction are shown in Fig.~\ref{slope}.
If the motion associated with the weak discontinuity were not
contributing to the sideward deflection, then the slope would
have been limited from above \cite{dan95} by $\chi/(2\sqrt{1 +
\chi})$ where
$\chi = (\langle E^{\perp } \rangle /2 -
\langle E^{\parallel} \rangle )/\langle E^{\parallel} \rangle$
and the energy components are from a $b=0$ reaction.
For protons from
400 MeV/nucleon reactions this limit is equal to~0.22, while
slopes given in Fig.~\ref{slope} reach values twice as high.

\section{Comparisons to FOPI Data}

As $b \rightarrow 0$, the expansion of matter
between the shocks, favoring the transverse direction of motion,
competes with the transparency effects in
the corona, that favor the longitudinal direction.
For~light systems in simulations, such as Ca~+~Ca,
the~transparency effects prevail, leading to
$\langle
E^{\perp} \rangle
/2 \langle E^{\parallel} \rangle < 1$
at $b = 0$, but not so
for heavy systems such as Au~+~Au, as was discussed and as is
evidenced in~Fig.~\ref{contours}.
The~experimental detection of
$\langle
E^{\perp} \rangle
/2 \langle E^{\parallel} \rangle > 1$, i.e.\
$\langle
E^{\perp} \rangle
/ \langle E^{\parallel} \rangle > 2$, would
demonstrate a~violence of the nuclear hydrodynamic
phenomena in heavy systems.
On the other hand, if for any reason
the~in-medium NN cross sections were lower than the free-space
cross-sections used in the simulations, then the~transparency
effects could
prevail over the anisotropy in the expansion, even within heavy
systems.  A~clear experimental answer requires a~full $2\pi$
coverage either in the forward or in the backward c.m.\
hemisphere.

\begin{figure}
  \vspace*{8.0cm}
  \includegraphics{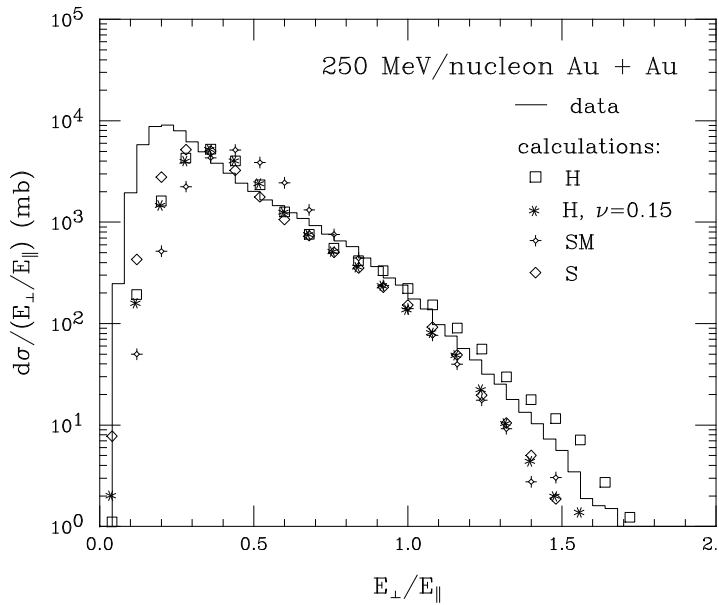}
\caption{
Differential $ERAT$ (cf.~Eq.~(\protect\ref{erat})) cross section
in Au + Au reaction at 250~MeV/nucleon as a~function of
$ERAT$.  The~data
of Refs.~\protect\cite{wie93,wie94,rei95} are represented by the
histogram.  The results of calculations utilizing
a~momentum-independent mean field, that yields a~stiff equation
of state, are represented by squares for free NN cross sections,
and by stars for density-dependent NN cross sections with $\nu = 0.15$,
cf.~Eq.~(\protect\ref{sigr}).  The results of calculations
utilizing free cross sections and mean fields, that yield
a~soft equation of state, are represented by crosses and
diamonds, respectively, for fields with and without
momentum dependence.  Final momentum distributions
in the calculations have been filtered through the experimental
acceptance.
 }
\label{ert250}
\end{figure}

Generally, the
experimental determination of the average anisotropy of
momentum distribution in $b \sim 0$ collisions may be
difficult, due
to the contaminations of event samples by events corresponding
to intermediate~$b$.  The idea borne within the FOPI
collaboration~\cite{rei95} has been then to employ, for
comparisons to model calculations, the differential cross
sections
for anisotropies quantified in terms of the {\em
event-by-event} energy ratio, sometimes termed~$ERAT$,
\begin{equation}
{E_{\perp} \over E_{\parallel}} = {\sum_{\nu} { p_{\nu}^{\perp
2} \over 2 m_{\nu}}
\over \sum_{\nu} { p_{\nu}^{\parallel 2} \over 2 m_{\nu}} },
\label{erat}
\end{equation}
where the sums extend over detected charged particles from
collisions emerging in the forward c.m.~hemisphere.  In
Phase~I, the FOPI detector setup covers laboratory angles
$1^{\circ} < \theta_{lab} < 30^{\circ}$ for the full azimuth,
with a broader range expected in Phase~II.
The~cross
sections for the energy ratio measured~\cite{wie93,wie94,rei95}
in
the 250 MeV/nucleon Au + Au reaction are shown in
Fig.~\ref{ert250}
together with the results of calculations filtered through the
acceptance of the FOPI apparatus.  The~calculations were
carried out for a~stiff ($H$) and for a~soft ($S$) equation of
state,
without ($H$ and $S$) and with ($SM$) momentum dependence, for
free-space cross sections and for cross sections reduced (in
the $H$ case) by a~factor exponentially dependent on
density,
\begin{equation}
\sigma_{NN}^{med} = \sigma_{NN} \exp(-\nu
\rho/\rho_0) \, .
\label{sigr}
\end{equation}

\begin{figure}
  \vspace*{8.0cm}
  \includegraphics{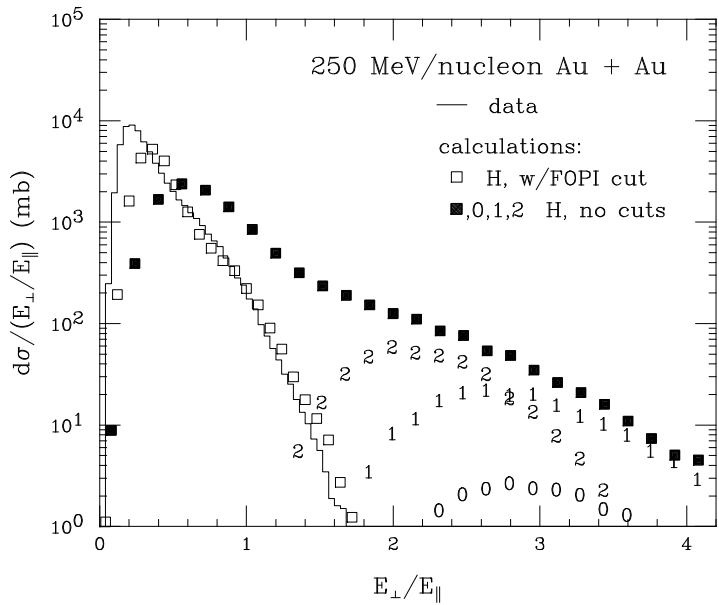}
\caption{
Differential $ERAT$ cross section
in Au + Au reaction at 250~MeV/nucleon as a~function of
$ERAT$.  The~data
are represented by the
histogram.  The results of an~H calculation utilizing free NN
cross-sections
are represented by open and filled squares
for the cases with and without experimental cuts, respectively.
  The symbols '0', '1', and '2',
represent contributions to the $ERAT$ cross section obtained
without experimental cuts, from
the impact-parameter intervals $0 < b < 0.5$ fm, 0.5 fm $<$ 1.5
fm, and 1.5 fm $<$ 2.5 fm, respectively.  All the results of
calculations were obtained using the forward c.m.~hemisphere only.
 }
\label{ert250f}
\end{figure}

\begin{figure}
  \vspace*{8.1cm}
  \includegraphics{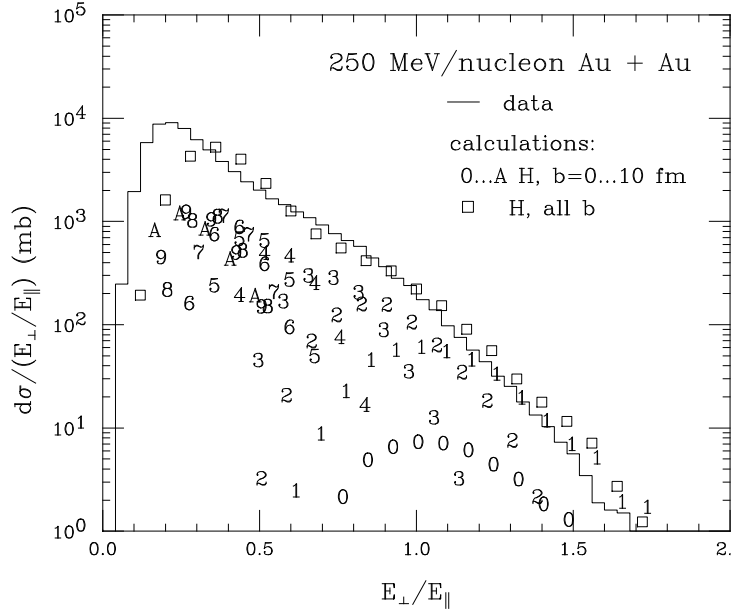}
\caption{
Differential $ERAT$ cross section
in Au + Au reaction at 250~MeV/nucleon as a~function of
$ERAT$.  The~data
are represented by the
histogram.  The results of an~H calculation
with free NN
cross-sections
are represented by squares.  The~symbols '0', '1',
'2',
\ldots, 'A', represent contributions to the calculated cross
section from impact parameter intervals
$0 < b < 0.5$ fm, 0.5 fm $<$ 1.5
fm, and 1.5 fm $<$ 2.5 fm, \ldots, 9.5 fm $<$ 10.5 fm,
respectively.
 }
\label{ert250b}
\end{figure}

Because of the acceptance cuts, no events above or in the
vicinity of
$E_{\perp} / E_{\parallel} = 2$ can be actually observed in
Fig.~\ref{ert250}.  The~comparison
between the data and calculations
at 250~MeV/nucleon and at other energies,
shows that the data allow only for a~small reduction ($\nu
\simeq 0.05-0.15$) in the NN cross-sections in
the simulations with a~stiff equation of
state and hardly for any reduction in the simulations with
a~soft equation.  Values as large as $\nu \gapproxeq 0.4$, for
the stiff
equation of state, are required for the transparency to prevail
over the transverse expansion in the $b=0$ Au + Au collisions.
When the
acceptance cuts are removed, the~$ERAT$ cross sections from
simulations extend well above $ERAT=2$, see Fig.~\ref{ert250f}.
It~should be possible to observe such distributions within the
Phase~II of the FOPI experiment.
Figure~\ref{ert250b}
shows the differential $ERAT$ cross section
computed with cuts and broken down into contributions from
narrow $b$-intervals.  The~figure demonstrates that selecting
the highest $ERAT$ values can yield a~surprisingly pure sample
of most central events with $b < 1.5$~fm.

\begin{figure}
\vspace*{10.8cm}
  \includegraphics{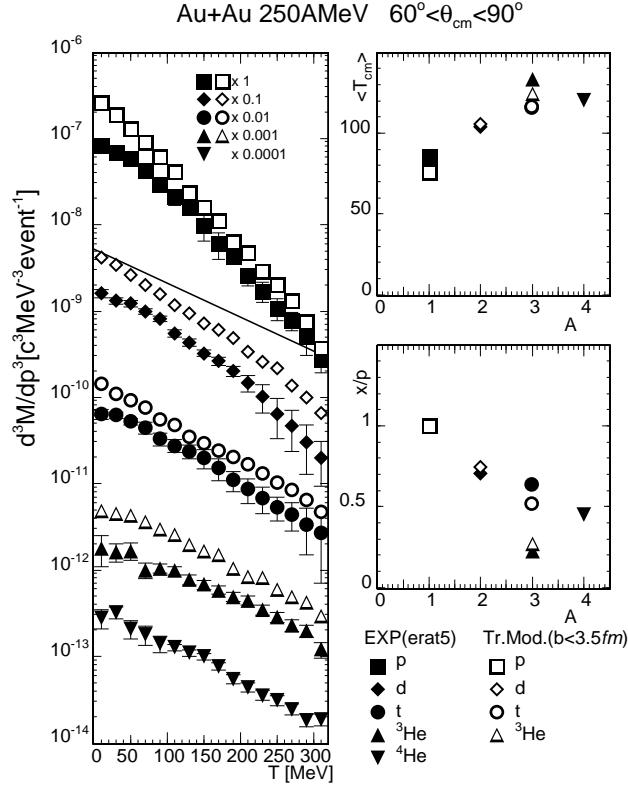}
\caption{Comparison of light-particle
measurements \protect\cite{pog95} (filled symbols) to
transport-model
calculations \protect\cite{dan95} (open symbols) for the central
Au
+ Au reaction at 250 MeV/nucleon.  Left, upper right, and lower
right panels show, respectively, transverse c.m.~spectra, mean
c.m.~energies, and transverse particle yields normalized to
proton yields.
Thin straight line across the c.m.~spectrum is drawn to guide
the eye, as parallel to the helion spectrum.
} \label{poggi}
\end{figure}

The transport simulations demonstrate various effects of the
anisotropy in the collective expansion on observables from the
collisions.  However, the~collective expansion should also
affect observables at any one angle.  Thus, in~a~globally
equilibrated
system, exponential spectra would be expected at any angle,
with
the same slope parameter for different particles, equal to the
temperature,
$d \, N_x/d \, p^3 \propto {\rm exp}(-E_x/T)$.  In~the
presence of the collective expansion, on the other hand,
the~particle spectra should exhibit different slopes for
particles with different mass, with particle spectrum becoming
flatter as particle mass increases.  This is due to
an~increased sensitivity to the collective motion with
an~increasing mass; for the average kinetic energy of
a~particle, one~would, in~fact, expect a~linear rise with the
mass number,
\begin{equation}
\langle E_x \rangle = {3 \over 2} \, T + {m_x \, \langle v^2
\rangle \over 2} =
{3 \over 2} \, T + A_x \, {m_N \, \langle v^2
\rangle \over 2} \, ,
\end{equation}
where $v$ is collective velocity and a~uniform temperature~$T$
is assumed.  The~spectra of light particles from the FOPI
measurements~\cite{pog95} and from the
calculations~\cite{dan95} are shown in Fig.~\ref{poggi},
together with the average energies as a~function of the mass
number and with the relative particle yields.
Similar flattening of the spectra with particle mass is
observed for the data and for the calculations.  The~average
kinetic energies are consistent and they rise approximately
linearly with the mass for the lightest fragments, exhibiting
the presence of the collective expansion.  Also the particle
yield ratios, shown in Fig.~\ref{poggi}, are rather consistent
at~250~MeV/nucleon between the theory and experiment.
These ratios reflect the entropy produced
within the system, as, in the limit of Boltzmann statistics,
\be
{x \over p} \propto \exp{\left(- A_x \, {S \over A} \right)} .
\label{xp}
\ee

\section{Entropy Production}

While the entropy produced in reactions attracted much
experimental attention in the past, this has not been the case
recently.  Yet, the entropy ties in an~interesting fashion to
the reaction dynamics and it may, principally, provide
information on the low-density phase transition.

\begin{figure}
  \vspace*{7.7cm}
  \includegraphics{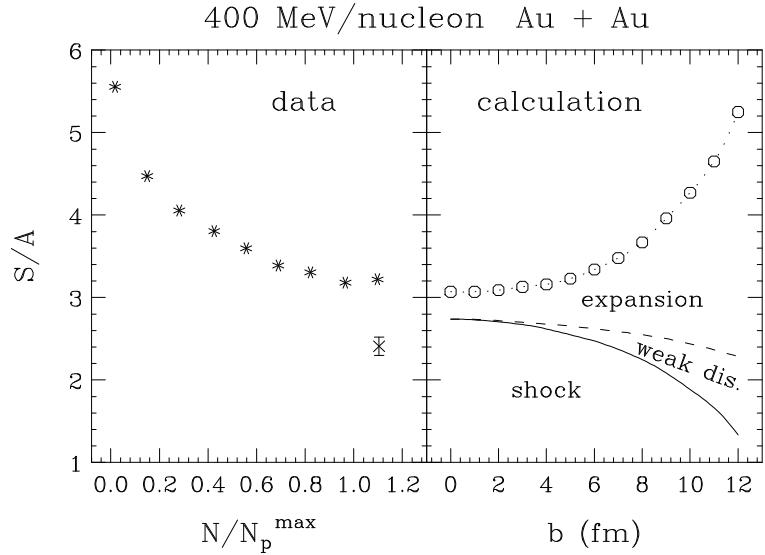}
\caption{
Left panel shows entropy per nucleon as a function of reduced
participant multiplicity, determined from
400~MeV/nucleon
Au~+~Au wide-angle light-particle data of
Ref.~\protect\cite{dos88a} (stars) and from intermediate-mass
fragment data of Ref.~\protect\cite{kuh93} (cross).
Right panel shows, as a function of impact parameter,
calculated~\protect\cite{dan95} total entropy per nucleon
associated with nucleons emitted
into wide angles (circles and dashed
line) and contributions to that entropy from  shock
waves
(solid line), from diffusion of vorticity around weak
discontinuity (between the solid and dashed lines), and from
 expansion of participant region (between the dashed and
dotted lines).  }

\label{entropy}
\end{figure}

The entropy in central reactions of heavy nuclei comes from
several
sources. Most of the entropy per nucleon in the participant
region at $b \sim 0$ is produced in shock waves,
see~Figs.~\ref{profau}, \ref{rankhu}, and~\ref{entropy}.
Additional entropy is produced during expansion.
The~hydrodynamic
Navier-Stokes equations yield for the rate of change of the
local entropy per baryon with time
\bea
\nonumber
\lefteqn{
{\partial \over \partial t} \left( {S \over A} \right) + {\bf
v} \cdot {\bf \nabla} \left( {S \over A} \right)
}  \hspace{2em} \\
& = & {1 \over
\rho T} \hbox{div} \left( \kappa {\bf \nabla} T \right)  +
{\eta
\over 2 \rho T} \left( {\partial v_i \over \partial r_k} +
{\partial v_k \over \partial r_i} - {2 \over 3} \, \delta_{ik}
\, {\partial v_l \over \partial r_l} \right)^2 ,
\label{enpro}
\eea
where $\kappa$ is heat capacity coefficient and $\eta$ is
viscosity coefficient.  Whereas the viscosity term on the rhs
of~(\ref{enpro}) always leads to the~rise in local entropy,
the~heat conduction term may cause a~local drop,
although still the entropy increases on a~global scale.
The~latter is
the case when the temperatures on the average decrease as one
moves away from a~considered point.

In the collective expansion, the~edges of participant matter
cool off; in an~isentropic expansion
$\rho T^{-3/2} = const$ and thus $T \rightarrow 0$ as
$\rho \rightarrow 0$.  Then the heat conduction, or simply flux
of energetic nucleons from the inner region, reheats
the outside matter, significantly raising the entropy there,
see Fig.~\ref{profau}.  Within the inner region the entropy per
nucleon decreases.  For $b \sim 0$ 400~MeV/nucleon
Au + Au collision, analytic estimates with Eq.~(\ref{enpro})
yield an~overall increase in the entropy $\delta(S/A)
\simeq 2  {\lambda
\over R}  \left( {\rho_{\rm comp} \over \rho_{\rm freeze}
} \right)^{5/6} \sim 1$
on account of the heat conduction, and, practically, no
increase
on account of viscosity due to the lack of shear in a~central
collision.  As~$b$ increases, shear appears and then the
entropy
gets to be produced through viscosity as the weak viscosity
spreads, cf.~Fig.~\ref{entropy}.  Moreover, the~production of
the entropy per nucleon in the participant region during
expansion rises, since the size of the region diminishes,
$\delta (S/A) \propto A^{-1/3}$.

\begin{figure}
  \vspace*{9.5cm}
  \includegraphics{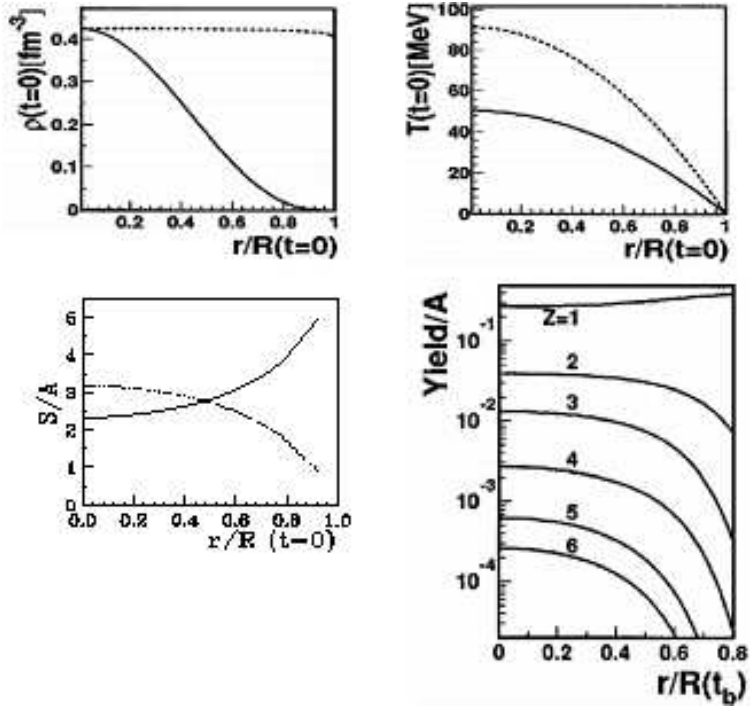}
\caption{
Initial fireball density (top left panel), temperature (top
right), and entropy per nucleon (bottom left), and final
fragment multiplicity (bottom right) as a~function of position
within the fireball.  Solid and dotted lines correspond to the
shape parameter~\protect\cite{pet95,bon78} $\alpha = 3.0$ and
$\alpha = 0.01$, respectively.
}

\label{petro1}
\end{figure}

\begin{figure}
  \vspace*{8.5cm}
  \includegraphics{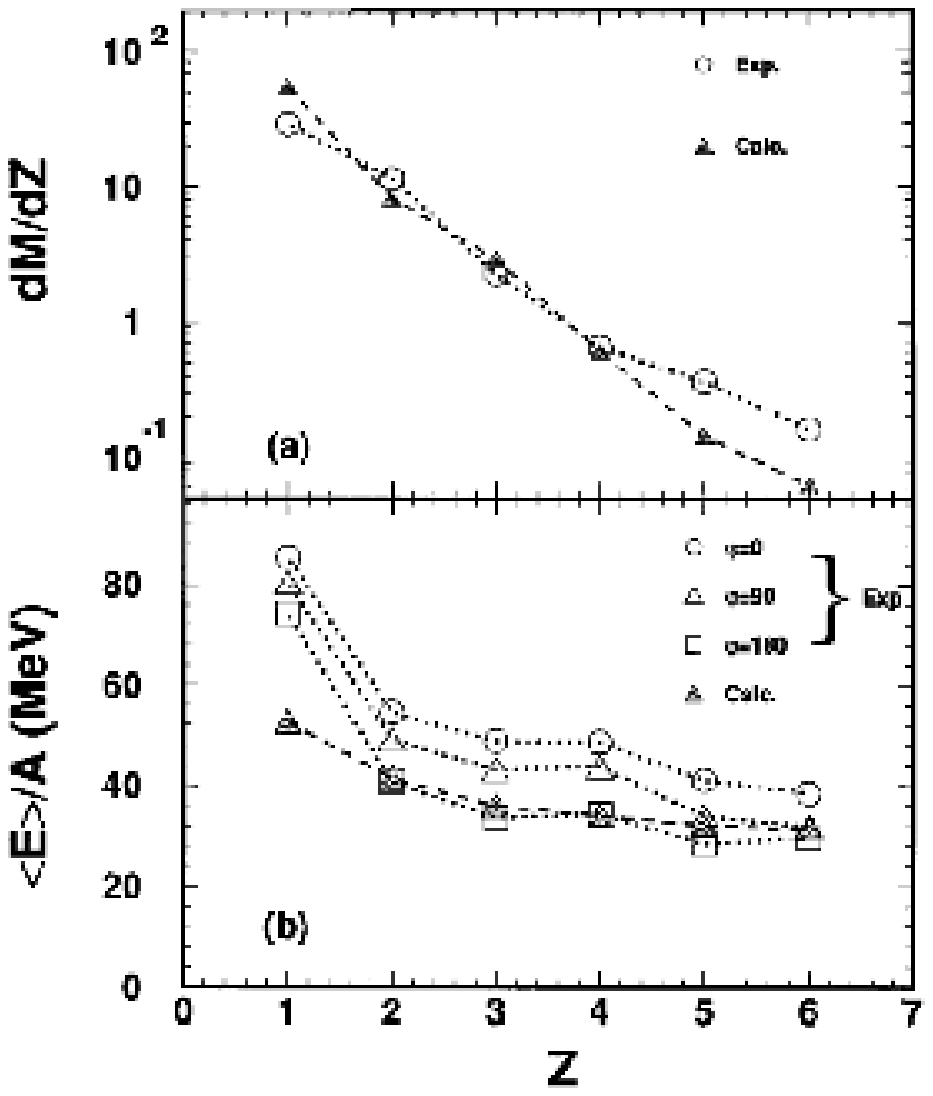}
\caption{
(a)~Average multiplicity in the forward hemisphere of central
250~MeV/nucleon Au + Au reaction~\protect\cite{pet95} and
(b)~average kinetic energy per nucleon for different azimuthal
intervals, as a~function of fragment charge.  Measurements are
represented by open symbols.  Results of the model
calculation for the optimal $\alpha = 0.03$ filtered
through the experimental acceptance are represented by filled
triangles.
}

\label{petro3}
\end{figure}

In a~recent investigation~\cite{pet95}, Petrovici \etal
\, attempted to describe, simultaneously, the~yields and the
average energies of intermediate-mass fragments from central
250~MeV/nucleon Au + Au collisions, in terms
of an~analytic ideal-fluid hydrodynamic model~\cite{bon78}
combined with a~model of statistical disassembly.
Within an~ideal-fluid model the entropy per nucleon is
conserved along a~streamline; there is no entropy production.
If~the initial entropy rises or declines with distance the from
the center, so~does the entropy towards the freeze-out.
The~parameter within the model, regulating the shape of initial
density, regulated simultaneously the entropy.  Best fit to
data was obtained for the entropy sharply rising with the
distance from the center, cf.~Figs.~\ref{petro1}
and~\ref{petro3},
consistent with Fig.~\ref{profau}.  Such
variation of the entropy favors, cf.~Eq.~(\ref{xp}),
the~production of heavier
fragments within the inner region of matter and of lighter
fragments outside, see Fig.~\ref{petro3}.  The~collective
energy per nucleon is lower inside and higher outside within
the model and according to the simulations.

As is apparent, the~change in entropy with the impact
parameter is tied to quite nontrivial aspects of reaction
dynamics.
Even at one impact parameter, in very central reactions,
fragments with different masses may be associated, due to
a~variation of the entropy, with different
regions and emission times.  It~may be worthwhile to study
directly the entropy and the collective energy as a~function of
mass accessing the spatial and temporal changes in a~reaction.

\section{Conclusions}

To summarize,  the dynamics of central energetic heavy-ion
collisions proceeds through numerous complex stages.  As~the
beam energy increases and/or the impact parameter decreases,
the maximum
compression increases.  Hydrodynamic behavior of matter
sets in the vicinity of balance energy.
At~that
energy, the~bulk pressure competes with viscous effects and
with the proximity force.  The~$\gamma$ polarization
measurements show directly the change in sign of the nuclear
deflection at the balance energy.  At~higher energies shock
fronts are observed to form in the head-on reaction
simulations,
perpendicular to beam axis and separating the hot compressed
matter from the cold.
In the semicentral
collisions a weak
tangential discontinuity develops in-between the shock fronts.
The hot compressed matter exposed to the vacuum in directions
parallel to the shock fronts  begins to expand collectively
into these directions.  The expansion affects particle angular
distributions and mean energy components and further shapes of
spectra and mean energies of particles emitted into any one
direction.  The~variation of slopes and the relative yields
measured
within the FOPI collaboration are in a~general agreement with
the results of simulations.
As~to the FOPI data on stopping, they
are consistent with the
preference for transverse over the longitudinal motion in the
head-on Au + Au
collisions.  However, the data cannot be used to
decide directly on that preference due to acceptance cuts.
Tied to the spatial and temporal changes in the reactions are
changes in the entropy per nucleon.  Different fragments are
associated with regions of different entropy values.  Besides
the participant entropy varies with impact parameter.
The~entropy sources include shock waves, weak discontinuity and
dissipation in expansion.

\section*{Acknowledgment}
This work was partially supported by the National Science
Foundation under Grant PHY-9403666.

\end{document}